\documentclass[aps,onecolumn,pra,superscriptaddress,showpacs,showkeys,floatfix]{revtex4-1}
\usepackage{graphicx,color}% Include figure files
\usepackage{epsfig}
\usepackage{graphicx}
\usepackage{mathrsfs}
\usepackage{color}
\usepackage{bm}
\usepackage{amsmath,amssymb,amsthm,amscd}
\usepackage{bbold}
\usepackage{mdwlist}

\newcommand{\Tr}{\operatorname{Tr}}

\newcommand{\dket}[1]{| \, #1 \rangle\!\rangle}
\newcommand{\dbra}[1]{\langle\!\langle #1 \, |}
\def\kk{\rangle\!\rangle}

\def\>{\rangle}
\def\<{\langle}

\DeclareMathOperator{\rank}{rank}

\allowdisplaybreaks

\begin{document}

\title{Mixed-state certification of quantum capacities for noisy communication channels}

\author{Chiara Macchiavello}
\affiliation{Quit group, Dipartimento di Fisica, 
Universit\`a di Pavia, via A. Bassi 6, 
 I-27100 Pavia, Italy}
\affiliation{Istituto Nazionale di Fisica Nucleare, Gruppo IV, via A. Bassi 6,
  I-27100 Pavia, Italy}

\author{Massimiliano F. Sacchi}
\affiliation{Istituto di Fotonica e Nanotecnologie - CNR, Piazza Leonardo
  da Vinci 32, I-20133, Milano, Italy}
\affiliation{Quit group, Dipartimento di Fisica, 
Universit\`a di Pavia, via A. Bassi 6, 
 I-27100 Pavia, Italy}
	
\date{\today}

\begin{abstract} 

We extend a recent method to detect lower bounds to the quantum
capacity of quantum communication channels by considering realistic
scenarios with general input probe states and arbitrary detection
procedures at the output.  Realistic certification relies on a new
bound for the coherent information of a quantum channel that can be
applied with arbitrary bipartite mixed input states and generalized
output measurements.

\end{abstract}

\maketitle 
The quantum capacity represents a central quantitative
notion in quantum information science \cite{lloyd,barnum,devetak,hay}.
However, in general, its computation is a hard task, since it requires
a regularisation procedure over an infinite number of channel uses,
and it is therefore by itself not directly accessible
experimentally. Its analytical value is known mainly for some channels
that have the property of degradability \cite{deveshor,ydh,rusk},
since regularisation is not needed in this case.  

\par In many practical situations a complete knowledge of the kind of
noise present along the channel is not available, and sometimes noise
can be completely unknown.  It is then important to develop efficient
means to establish whether in these situations the channel can still
be profitably employed for information transmission.  A standard
method to establish this relies on quantum process tomography
\cite{nielsen97,pcz,mls,dlp,alt,cnot,vibr, ion,mohseni,irene,atom},
where a complete reconstruction of the completely positive map
describing the action of the channel can be achieved, and therefore
all its communication properties can be estimated. This, however, is a
demanding procedure in terms of the number of different measurement
settings needed, since it scales as $d^4$ for a finite $d$-dimensional
quantum system.  

\par When one is not interested in reconstructing the complete form of
the noise affecting the channel but only in detecting its quantum
capacity, which is a very specific feature, a novel and less demanding
procedure in terms of needed resources (measurements) has been
presented in Ref. \cite{ms16}.  In the same spirit as it is done, for
example, in entanglement detection \cite{ent-wit}, parameter
estimation \cite{parest}, and detection of entanglement-breaking
property \cite{qchanndet} or non-Markovianity \cite{nomadet} of
quantum channels, the method of Ref. \cite{ms16} allows one to
experimentally detect lower bounds to the quantum capacity by means of
a number of local measurements that scales as $d^2$. The method can be
applied to generally unknown noisy channels, and has been proved to be
very efficient for many examples of single qubit channels, for
generalized Pauli channels in arbitrary dimension \cite{ms16}, and for
two-qubit memory Pauli and amplitude damping channels
\cite{ms-corr}. The first experimental demonstration has been also
recently shown in Ref. \cite{exp}, based on a quantum optical
implementation for various forms of noisy single-qubit channels,
proving the feasibility and efficiency of the method.

\par In the original proposal of Ref. \cite{ms16} we considered a pure
maximally entangled input state as a probe, which was used to sample
the channel and reconstruct the probabilities for output measurements
over orthogonal projectors.  In this paper we extend our certification
method by considering the case of a generally mixed bipartite input
state and probabilities pertaining to generalized measurements
(i.e. POVM's).  Clearly, such a generalization makes the method more
flexible {\em per se}, and also allows one to compare theoretical
predictions with experiments, where pure states and perfect
measurements always represent an idealization.  In fact, already in
Ref.  \cite{exp}, a specific treatment of the experimental data was in
order, since the ideal input maximally entangled state was
realistically replaced with a Werner state, because of unavoidable
imperfections in the procedures of quantum-state preparation.

Let us denote the action of a 
generic quantum memoryless channel on a single system as ${\cal E}$ and define 
${\cal E}_N= {\cal E}^{\otimes N}$, where $N$ represents the number of 
channel uses.
The quantum capacity $Q$ measured in qubits per channel use is defined 
as \cite{lloyd,barnum,devetak,hay}
\begin{eqnarray} Q=\lim _{N\to \infty}\frac
{Q_N}{N}\;,\label{qn} 
\end{eqnarray} 
where
$Q_N = \max
_{\rho } I_c (\rho , {\cal E}_N)$, 
and $I_c(\rho , {\cal E}_N)$ denotes the coherent information 
\cite{schumachernielsen}
\begin{eqnarray} I_c(\rho , {\cal E}_N) = S[{\cal E}_N (\rho )] - S_e
(\rho, {\cal E}_N)\;.\label{ic} \end{eqnarray} 
In Eq. (\ref{ic}),
$S(\rho )=-\Tr [\rho \log _2 \rho ]$ is the von Neumann entropy, and
$S_e (\rho, {\cal E})$ represents the entropy exchange \cite{schumacher}, i.e.
$S_e (\rho, {\cal E})= S[({\cal I}_R  \otimes
{\cal
E} 
)(|\Psi _\rho \rangle \langle \Psi _\rho |)] $,
where $|\Psi _\rho \rangle $ is any purification of $\rho $ by means of a 
reference quantum system $R$, namely 
$\rho =\Tr _R [|\Psi _\rho \rangle \langle \Psi _\rho|]$.

\par Let us consider an arbitrary bipartite input mixed state $\sigma $ 
for the tensor-product Hilbert space 
of reference and system, and write it as a convex decomposition (not necessarily the  spectral one) 
of pure states, namely  
\begin{eqnarray}
\sigma =\sum _l a_l \dket{A_l}\dbra{A_l} \;,\label{simi}
\end{eqnarray}
with $a_l \geq 0$ and $\sum _l a_l=1$. The double-ket notation $\dket{A}$ introduced in Eq. (\ref{simi})   
is useful to remind one of the isomorphism between bipartite vectors 
$\dket{A}=\sum_{n,m}A_{nm} |n 
\rangle |m \rangle $ and linear operators $A=\sum_{n,m}A_{nm} |n 
\rangle \langle m |$, along with the identities \cite{pla} 
\begin{eqnarray}
\dbra{A} B \kk =\Tr [A^\dag B]\;,
\end{eqnarray}
and 
\begin{eqnarray}
A \otimes B \dket{C} = \dket{AC B^\tau }\;,
\end{eqnarray}
where $\tau $ denotes transposition on a fixed basis. 
The partial trace of $\sigma  $ over the reference is the system mixed state
\begin{eqnarray}
\rho = \Tr_{R}[\sigma ]=\left (\sum _l a_l A_l ^\dag  A_l \right )^{\tau }
\;.\label{trr}
\end{eqnarray}
Since any purification of a mixed state $\rho $ of the system can be written as 
$\dket{V \sqrt{\rho ^\tau }} $
with arbitrary unitary $V$, then  
the entropy exchange for $\rho $ is given by 
\begin{eqnarray}
S_e(\rho, {\cal E})
= S[({\cal I}_R \otimes {\cal E}
) 
(\dket{V \sqrt{\rho ^\tau }} \dbra{V \sqrt{\rho ^\tau }})]
\;.
\end{eqnarray}
By writing the following spectral decomposition 
\begin{eqnarray}
({\cal I}_R \otimes {\cal E}) 
(\dket{V \sqrt{\rho ^\tau }} \dbra{V \sqrt{\rho ^\tau }})]
\equiv \sum _j s_j \dket{\phi _j} \dbra{\phi _j}
\;,
\end{eqnarray}
one has 
\begin{eqnarray}
S_e(\rho, {\cal E}) 
= - \sum _j s_j \log_2 s_j\;. 
\end{eqnarray}
For full-rank $\rho = \Tr_{R}[\sigma ]$, 
the probability $p_i$ for input mixed state $\sigma $ as in Eq. (\ref{simi}) and 
measurement outcome $i$ pertaining to the element $\Pi _i$ of an arbitrary POVM 
for the tensor product of reference and system
is given by 
\begin{align}
p_i & = \Tr [
\sigma  \Pi_i ]
%=\nonumber \\& & 
=\sum _l a_l \Tr [
({\cal I}_R 
\otimes {\cal E}) (\dket{I}\dbra{I}) (A^\dag _l \otimes I) \Pi_i (A_l \otimes I)] 
\nonumber \\&  
=\sum _l a_l \Tr \left [({\cal I}_R 
\otimes {\cal E}) 
(\dket{V\sqrt {\rho ^\tau }}
\dbra{V \sqrt {\rho ^\tau }}) \left ( V \frac {1}{\sqrt {\rho ^\tau }} 
A^\dag _l \otimes I \right ) \Pi_i \left (A_l \frac {1}{\sqrt {\rho ^\tau }} V^\dag \otimes I \right )
\right ] 
\nonumber \\&  =
\sum_j s_j \dbra{\phi _j} \sum _l a_l  
\left ( V \frac {1}{\sqrt {\rho ^\tau }} 
A^\dag _l \otimes I \right ) \Pi_i \left (A_l \frac {1}{\sqrt {\rho ^\tau }} V^\dag \otimes I \right )
\dket{\phi _j}
\nonumber \\&  
\equiv \sum _{j} s_j p(i|j)
\;,\label{pimixed}
\end{align}
where in the last line we introduced the conditional probability 
\begin{eqnarray}
p(i|j)= \dbra{\phi _j} \sum _l a_l  
\left ( V \frac {1}{\sqrt {\rho ^\tau }} 
A^\dag _l \otimes I \right ) \Pi_i \left (A_l \frac {1}{\sqrt {\rho ^\tau }} V^\dag \otimes I \right )
\dket{\phi _j}
\;.
\end{eqnarray}
By denoting 
the Shannon entropy for the vector of the 
probabilities $\{p_i\}$ as $H(\vec p)= -\sum_{i} p_i \log_2 p_i$, then one has
\begin{eqnarray}
&&S_e(\rho, {\cal E}) - H(\vec p)=
\sum _i p_i \log _2 p_i -\sum _j s_j \log _2 s_j  \nonumber \\& & = 
\sum _{i,j} s_j p(i|j)\log _2 \frac {p_i}{s_j} 
\leq  \log _2 \left (
\sum _{i,j} s_j p(i|j) \frac{p_i}{s_j} \right ) \nonumber \\& & =
\log _2  \vec r \cdot \vec p  \leq \log  _2  \vec t \cdot \vec p
\;,\label{dife}
\end{eqnarray}
where we used Jensen's inequality in the second line,  
and defined  the vectors $\vec r$ and 
$\vec t$ with components $r_i = \sum _{j} p(i|j)$ and 
\begin{eqnarray}
t_i = \Tr \left [\left 
(\sum _l a_l A_l \frac {1}{\rho ^\tau } A^\dag \otimes I \right ) \Pi _i \right ] \geq r_i\;,
\label{timix}
\end{eqnarray}
respectively. Notice that the elements of $\vec t$ are independent of the unknown channel 
and can be evaluated from the explicit form of the input state and the output measurement. 
For a mixed state $\rho $ in Eq. (\ref{trr}) which is not full-rank, since 
$\ker \rho ^\tau  = \cap _l A_l$ and hence $\ker \rho ^\tau \subset \ker A_l$ for all $l$, 
it is easy to see that Eq. 
(\ref{timix}) is replaced with 
\begin{eqnarray}
t_i =\Tr \left [\left (\sum _l a_l A_l (\rho ^\tau )^+ A^\dag _l \otimes I \right )\Pi _i \right ]
\;,\label{timor}
\end{eqnarray}
where $M^+$ denotes the Moore-Penrose pseudo inverse of $M$ \cite{moore}.   
Eq. (\ref{dife}) provides the following bound for the entropy exchange 
\begin{eqnarray}
S_e\left (\rho , {\cal E} \right ) \leq H (\vec p)+ 
\log  _2  \vec t \cdot \vec p\;,
\label{se-bound2}
\end{eqnarray}
where $p_i $ represents the probability of measurement outcome in Eq. (\ref{pimixed}) 
for the mixed input state (\ref{simi}), and the components $t_i$ are given in Eq. (\ref{timor}). 
From Eqs. (\ref{qn}), (\ref{ic}), and (\ref{se-bound2}) it follows that for any $\rho$ and $\vec p$ one has 
the following chain of bounds 
\begin{eqnarray}
Q \geq Q_1 \geq I_c(\rho , {\cal E}_1)\geq S\left [{\cal E} (\rho )\right ]-H(\vec p)- \log  _2  \vec t \cdot \vec p 
\equiv Q_{DET}
\;.\label{qvec}
\end{eqnarray}

\par The lower bound $Q_{DET}$ to the quantum capacity of any unknown
channel can then be easily accessed without requiring full process
tomography of the quantum channel, by means of the following
procedure: $i)$ prepare a bipartite state $\sigma $ 
and send it through the unknown channel ${\cal I} _R \otimes {\cal E}$, where 
${\cal E}$ acts on one of the two subsystems;
$ii)$ measure suitable local observables on the joint output state to
estimate $\vec p$ and $S\left [{\cal E} (\rho )\right ]$ in order to
compute $Q_{DET}$; $iii)$ after performing the measurements, the
detected bound $Q_{DET}$ can be further optimized over all probability
vectors that can be obtained from the used measurement setting.
In fact, for a fixed measurement setting, one can infer different vectors 
of probabilities pertaining to different  POVM's $\{ \Pi _i \}$.  
This last step is achieved by performing ordinary classical
processing of the measurement outcomes. The bound for quantum capacity certification in Eq. (\ref{qvec}) 
generalizes the result of Ref. \cite{ms16}, where only a maximally entangled pure input state and set of 
orthogonal projectors were considered (for which $\log _2 \vec t \cdot \vec p =0$).

Differently from a complete process tomography, we do not need to
measure a complete set of observables and, moreover, the bound is
directly obtained from the measured expectations, without need of
linear inversion and/or maximum likelihood technique. Notice also
that, like in quantum process tomography assisted by an ancilla,
entanglement is not mandatory, since the bipartite input state $\sigma $ just
has to be faithful \cite{dlp,faith2}, namely such that the output
state $({\cal I}_R \otimes {\cal E}) (\sigma )$ is in one-to-one
correspondence to the map ${\cal E}$. 

Finally, we remark that the detectable bound (\ref{qvec}) 
also gives a lower bound to the private information $P$,  
since $P \geq Q_1$ \cite{NC00}, 
and to the  entanglement-assisted classical capacity $C_E$, since 
$C_E = \max_{\rho } [S(\rho ) + I_c (\rho, {\cal E}_1)]$, and then clearly 
$C_E \geq S(\rho ) + Q_{DET}$.

A number of particular cases for Eq. (\ref{timor}) can be inspected as follows: 
\begin{itemize}
\item[i)] 
If $\Pi _i$ is a projector on a maximally entangled state, namely $\Pi _i = \frac 1d 
\dket{U_i} \dbra{U_i}$ 
with $U_i$ unitary operator, then $t_i= \frac{\rank \rho}{d}$. 
\item [ii)] If the bipartite input state $\sigma $ is pure with maximal Schmidt number, namely 
$\sigma =\dket{A} \dbra{A}$  with $A$ invertible, then $t_i=\Tr [\Pi_ i]$.  
\item [iii)] If the reduced input state is $\rho = \frac I d$, then $t_i= d \Tr [ \Tr _S [\sigma ] 
\Tr _S [\Pi _i]]$. 
\item [iv)] If the input state $\sigma $ is diagonal on maximally entangled states, namely $A_l = \frac {U_l}{\sqrt d}$ 
where $\{U_ l\}$ is an operator basis of unitary operators, then $t_i=\Tr [\Pi_ i]$. 
\item [v)] Finally, when $t_i=\Tr [\Pi_ i]=k$, $k$ being a constant for all $i$ (and then 
necessarily $k=d^2/N$ for a POVM with $N$ elements), one has 
$\log _2 \vec t \cdot \vec p = \log _2 k$.      
\end{itemize}
Notice also that the relation $\sum _i t_i =d \rank \rho $ always holds. 

In the following we provide two examples of quantum capacity detection for specific 
bipartite mixed input states $\sigma  $ and channels $\cal E$, comparing the results with those 
of the original procedure of Ref. \cite{ms16},  where only a maximally entangled input state 
was considered.

{\em Example 1}.  
Let us consider a Pauli channel in dimension $d$ 
\begin{eqnarray}
 {\cal E}(\rho )=\sum _{m,n=0}^{d-1} p_{m,n} U_{mn} \rho U^{\dag }_{mn}\;,\label{pau}
\end{eqnarray}
where $U_{mn}$ represents the unitary operator 
$U_{mn}=\sum _{k=0}^{d-1} e^{\frac{2\pi i}{d} km} |k \rangle 
\langle (k + n)\!\!\!\mod d |$, and $\sum _{m,n=0}^{d-1} p_{m,n}= 1$, along with 
a bipartite mixed input state diagonal on the generalized 
Bell basis, namely 
\begin{eqnarray}
\sigma = \frac 1d \sum _{m,n=0}^{d-1} q_{m,n} 
\dket {U_{mn}} \dbra {U_{mn}}\;,\label{smn}
\end{eqnarray}
with $q_{m,n}\geq 0$, and $\sum _{m,n=0}^{d-1} q_{m,n} = 1$.
 
Since the generalized Bell projectors can be written as follows 
\cite{pla,bellobs}
\begin{eqnarray}
\Pi _{mn}= \frac 1d \dket{U _{mn}}\dbra{U _{mn}}=\frac {1} {d ^2} \sum _{p,q=0} ^{d-1} 
e^{\frac {2\pi i}{d}(np-mp) }U_{pq} \otimes U^*_{pq} 
\,,\label{umnbel}
\end{eqnarray}
with $*$ denoting complex conjugation, 
then a set of local  measurements on the eigenstates of $U_{mn} \otimes U^*_{mn}$ 
allows one to estimate $Q_{DET}$ in Eq. (\ref{qvec}), and    
one has 
\begin{eqnarray}
Q \geq Q_{DET}=\log _2 d - H(\vec p \, ') \;, 
\end{eqnarray}
where $\vec p \,'$ is the $d^2$-dimensional vector of probabilities 
pertaining to the generalized Bell projectors (\ref{umnbel}), 
whose components are given by 
\begin{eqnarray}
p_{m,n}' 
= \sum_{l,s=0}^{d-1} p_{l,s} q_{m-l,n+s}
\;.\label{pp}
\end{eqnarray}
For a pure maximally entangled state, e.g. $\sigma = \frac 1 d {\dket{I} \dbra{I}}$, 
one recover $p_{m,n}' = p_{m,n}$ for all $m,n$. 
In Ref. \cite {exp} one can find some experimental 
results for the qubit case and a number of different channels 
(dephasing, depolarizing, and Pauli), with a 
mixed input state of the specific form 
\begin{eqnarray} 
\sigma  
=\frac {4 F -1}{3} |\Phi ^+ \rangle \langle \Phi ^+| + \frac
{1-F}{3}I\otimes I 
\;,\label{iso} 
\end{eqnarray} 
namely a maximally entangled state $|\Phi ^+ \rangle $ affected by isotropic noise that 
reduces the fidelity 
from $1$ to $F=\langle \Phi^+ | \sigma  | \Phi^+ \rangle $. 

For the special case of a depolarizing channel 
[i.e. $p_{0,0}=1-p$ and $p_{m,n}=\frac{p}{d^2-1}$ for $ (m,n) \neq (0,0)$ in Eq. (\ref{pau})], 
by using an ideal maximally entangled input state $\frac{1}{\sqrt d}\dket{I}$, the 
detectable bound coincides with the hashing bound \cite{ms16}, 
namely 
\begin{eqnarray}
Q_{DET}=\log_2 d - H_2 (p)-p \log_2 (d^2-1)\;.\label{hash}
\end{eqnarray}
Let us consider now an isotropic input state for dimension $d$
\begin{eqnarray} 
\sigma  
= \frac
{d^2 F -1}{d^2 -1} \frac {\dket{I} \dbra{I}}{d}
+ \frac {1-F}{d^2
-1}I\otimes I \;,\label{iso2} 
\end{eqnarray} 
where $F$ represents the fidelity of $\sigma $ with the ideal maximally entangled state 
$\frac {1}{\sqrt d}\dket{I}$, Notice that such a choice corresponds to Eq. (\ref{smn}), 
with $q_{0,0}=F$ and $q_{m,n}=\frac{1-F}{d^2 -1}$  for $(m,n)\neq (0,0)$. By applying 
Eq. (\ref{pp}) one obtains 
\begin{eqnarray}
&&p'_{0,0}=(1-p)F +\frac{p(1-F)}{d^2-1},\nonumber \\& & 
p'_{m,n}=\frac{1}{d^2-1}(1-p'_{0,0}) \qquad \hbox{for }(m,n)\neq (0,0). 
\;
\end{eqnarray}
Hence, the detected quantum capacity in Eq. (\ref{hash}) is replaced with 
\begin{eqnarray}
Q_{DET}=\log_2 d - H_2 (p')-p'\log_2 (d^2-1)\;,\label{hash2}
\end{eqnarray}
where 
\begin{eqnarray}
p'=1-p'_{0,0}=\frac{d^2[1-F(1-p)]+F-p-1}{d^2-1}\;.
\end{eqnarray}
\begin{figure}[htb]
  \includegraphics[scale=0.8]{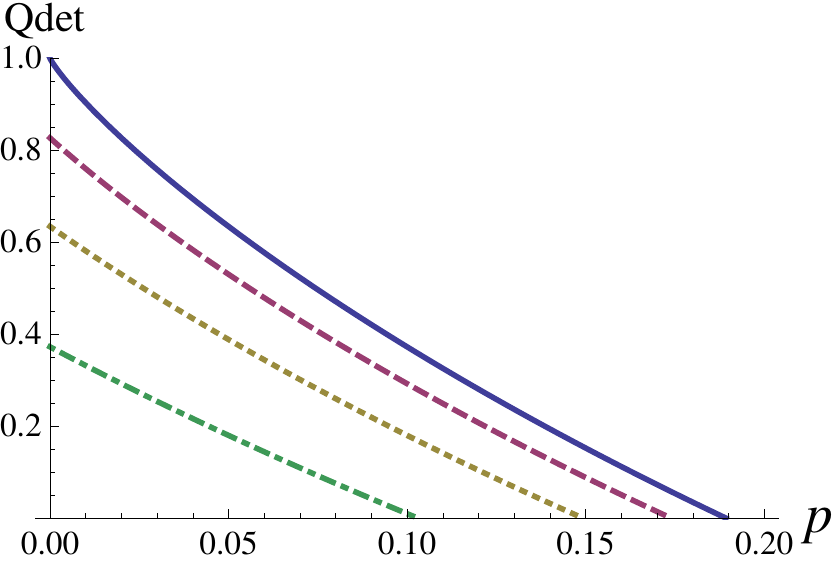}
  \caption{Detectable quantum capacity for the qubit depolarizing channel
  versus $p$ for mixed input state as in
  Eq. (\ref{iso2}) with $d=2$ and fidelity $F=0.98$ (dashed), $F=0.95$ (dotted), 
$F=0.9$ (dot-dashed), along
  with the hashing bound (solid line) achieved for $F=1$.}  
\end{figure}

In Fig. 1 
we plot the detectable quantum capacity for the qubit depolarizing channel 
for different values of the fidelity $F$ of the input state (\ref{iso2}). 
Clearly, for decreasing values of $F$, the certification of quantum capacity deteriorates. 
In fact, the ideal hashing bound (\ref{hash}) approaches zero for $p\gtrsim 0.1892$ \cite{div}. 
This threshold value of $p$ for certifying positive quantum capacity decreases when using noisy 
input states.  
We notice that for any value of $p$, if $F \lesssim 0.818$ 
no quantum-capacity certification is obtained since one has $Q_{DET}<0$.

{\em Example 2}. Erasure channel with erasure probability $p$ in dimension $d$, namely 
\begin{eqnarray}
{\cal E}(\rho )= (1-p) \rho \oplus p |e \rangle \langle e | \Tr [\rho ] \;, 
\end{eqnarray}
where $|e \rangle $ denotes the erasure flag which is orthogonal to the system Hilbert space. 
Since it is a degradable channel \cite{deveshor}, its quantum capacity coincides 
with the one-shot single-letter 
quantum capacity $Q_1$, and one has
\begin{eqnarray}
Q=Q_1=(1-2p)\log _2 d
\;,\label{qcerase}
\end{eqnarray}
for $p\leq \frac 12$, and $Q=0$ for $p\geq \frac 12$.
Let us consider a bipartite mixed input state as in Eq. (\ref{iso2}).  
The bipartite output is given by 
\begin{eqnarray}
({\cal I}_R\otimes {\cal E}) \sigma  =(1-p)\left[
\frac {d^2 F -1}{d^2 -1} 
\frac {\dket{I} \dbra{I}}{d} + \frac {1-F}{d^2 -1}I\otimes I \right] 
\oplus  \ p \left [\frac{I_R}{d}\otimes |e \rangle \langle e |  \right ]\;. 
\end{eqnarray}
A basis constructed by the union of the 
projectors on $|i \rangle \otimes |e \rangle  $ 
(with $i=0,1,\cdots, d-1$) and Bell projectors 
(where one of them corresponds to  $\frac 1d  \dket{I} \dbra{I}$) 
gives a vector of probability $\vec p$ with $d$ elements equal to
$p/d$  [corresponding to $|i \rangle \otimes |e \rangle  $],  one element equal to $(1-p)F$ 
[corresponding to $\frac 1d \dket{I} \dbra{I}$], and $d^2 -1$ elements equal to $(1-p)(1-F)/(d^2-1)$ 
[corresponding to projectors on maximally entangled states orthogonal to 
$\frac {1}{ \sqrt d} \dket{I} $]. 

We then have 
\begin{eqnarray}
H(\vec p)= H_2(p) +(1-p)H_2(F)+p\log_2 d +(1-p)(1-F)\log _2 (d^2 -1)
\;,  
\end{eqnarray}
where $H_2(p) \equiv -p \log_2 p - (1-p)\log _2 (1-p)$  denotes the binary Shannon entropy. 
The von Neumann entropy of the reduced output state 
${\cal E}\left (\frac Id \right )
= (1-p) \frac Id \oplus p 
|e \rangle \langle e | $ is given by 
$S \left [{\cal E} \left ( \frac Id \right )
\right ]
=H_2 (p)+(1-p)\log_2 d$.
Then, it follows that the detectable bound $Q_{DET}$ for the quantum capacity of the erasure channel 
is given by 

\begin{eqnarray}
Q \geq Q_{DET} \equiv (1-2p)\log_2 d - (1-p)[H_2(F)+(1-F)\log _2 (d^2 -1)]
\;.
\end{eqnarray}
Notice that for perfect fidelity $F=1$, one achieves $Q=Q_{DET}$. 

In Fig. 2 we compare the detectable quantum capacity for the qubit
erasure channel for different values of fidelity $F$ with the exact
quantum capacity.  We also remark that for any value of erasure
probability $p$, if $F \lesssim 0.811$ no quantum-capacity
certification is obtained since one has $Q_{DET}<0$.

\begin{figure}[htb]
  \includegraphics[scale=0.8]{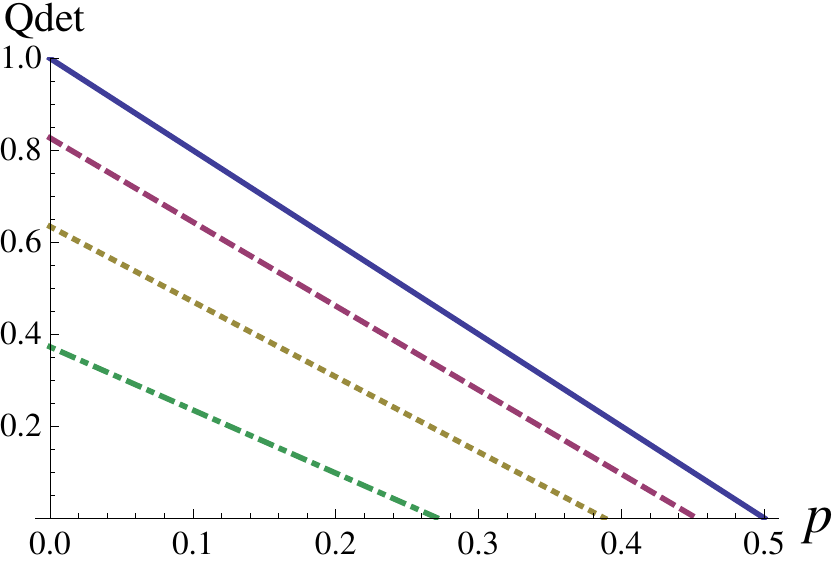}
  \caption{Detectable quantum capacity for the qubit erasure channel
  versus erasure probability $p$ for mixed input state as in
  Eq. (\ref{iso2}) with $d=2$ and fidelity $F=0.98$ (dashed), $F=0.95$ (dotted), 
$F=0.9$ (dot-dashed), along
  with the exact quantum capacity (solid line) achieved for $F=1$.}  \end{figure}

In conclusion, we have extended our recent  method to detect lower bounds to capacities 
of quantum communication channels (specifically, to the quantum capacity, the 
private capacity, and the entanglement-assisted classical capacity). 
This capacity certification does not require any
{\em a priori} knowledge about the quantum channel and relies on a number of measurement 
settings that scales as $d^2$, thus much more favorably than complete process
tomography. We think that the presented more general approach can be relevant for the 
realistic scenario where experimental imperfections 
are taken into account. 
In this way one can theoretically predict and evaluate the robustness of quantum-capacity 
witnessing 
with respect to input noisy states and output generalized measurements.


\begin{thebibliography}{99}


\bibitem{lloyd}
S. Lloyd, Phys. Rev. A \textbf{55}, 1613 (1997).

\bibitem{barnum}
H. Barnum, M. A. Nielsen, and B. Schumacher, 
Phys. Rev. A \textbf{57}, 4153 (1998).

\bibitem{devetak} I. Devetak, IEEE Trans. Inf. Theory \textbf{51}, 44 (2003).

\bibitem{hay} P. Hayden, M. Horodecki, A. Winter, and J. Yard, Open Sys. Inf. Dyn. {\bf 15}, 
7 (2008). 

\bibitem{deveshor} I. Devetak and P. Shor, Comm. Math. Phys.  
\textbf{256}, 287 (2005).

\bibitem{ydh} J. Yard, I. Devetak, and P. Hayden, IEEE Trans. Inf. Theory {\bf 54}, 3091 (2008).

\bibitem{rusk} T. S. Cubitt, M. Ruskai, and G. Smith, J. Math. Phys. {\bf 49}, 102104 (2008).

\bibitem{nielsen97} I. L. Chuang and M. A. Nielsen, Journal of Modern Optics \textbf{44}, 2455 (1997).

\bibitem{pcz} J. F. Poyatos, J. I. Cirac, and P. Zoller, Phys. Rev. Lett. {\bf 78}, 390 (1997).

\bibitem{mls}M. F. Sacchi, Phys. Rev. A {\bf 63}, 054104 (2001).

\bibitem{dlp} G. M. D'Ariano and P. Lo Presti, Phys. Rev. Lett. {\bf 86}, 4195 (2001). 

\bibitem{alt} J. Altepeter, D. Branning, E. Jeffrey, T. Wei, P. Kwiat, R. Thew, J. OBrien, M. Nielsen, and A. White, 
Phys. Rev. Lett. {\bf 90}, 193601 (2003).

\bibitem{cnot} J. L. O'Brien, G. J. Pryde, A. Gilchrist, D. F. V. James, N. K. Langford, T. C. Ralph, and A. G. White,
Phys. Rev. Lett. {\bf 93}, 080502 (2004).

\bibitem{vibr}S. H. Myrskog, J. K. Fox, M. W. Mitchell, and A. M. Steinberg, 
Phys. Rev. A {\bf 72}, 013615 (2005). 

\bibitem{ion}M. Riebe, K. Kim, P. Schindler, T. Monz, P. O. Schmidt, T. K. K\"orber, W. H\"ansel, H. H\"affner, 
C. F. Roos, and R. Blatt, 
Phys. Rev. Lett. {\bf 97}, 220407 (2006).

\bibitem{mohseni} M. Mohseni, A. T. Rezakhani, and D. A. Lidar, Phys. Rev. A \textbf{77}, 032322 (2008).

\bibitem{irene} I. Bongioanni, L. Sansoni, F. Sciarrino, G. Vallone, and P. Mataloni, Phys. Rev. A \textbf{82}, 042307 (2010).

\bibitem{atom}Y. Sagi, I. Almog, and N. Davidson, 
Phys. Rev. Lett. {\bf 105}, 053201 (2010).


\bibitem{ms16} C. Macchiavello and M. F. Sacchi, 
Phys. Rev. Lett. {\bf 116}, 140501 (2016). 

\bibitem{ent-wit} O. G\"uhne, P. Hyllus, D. Bru\ss, A.~Ekert,
M. Lewenstein, C.~Macchiavello, and A. Sanpera,
Phys. Rev. A {\bf 66}, 062305 (2002).

\bibitem{parest}
G. M. D'Ariano, M. G. A. Paris, and M. F. Sacchi, 
Phys. Rev. A {\bf 62}, 023815 (2000).

\bibitem{qchanndet} C. Macchiavello and M. Rossi, Phys. Rev. A  {\bf 88}, 
042335 (2013); A. Orieux, L. Sansoni, M. Persechino, P. Mataloni, M. Rossi,  
and C. Macchiavello, Phys. Rev. Lett. {\bf 111}, 220501 (2013).

\bibitem{nomadet} D. Chruscinski, C. Macchiavello, and S. Maniscalco, Phys. Rev. Lett. {\bf 118}, 080404 (2017).


\bibitem{ms-corr} C. Macchiavello and M. F. Sacchi, 
Phys. Rev. A \textbf{94}, 052333 (2016).

\bibitem{exp} A. Cuevas, M. Proietti, M. A. Ciampini, S. Duranti, P. Mataloni, M. F. Sacchi,  
and C. Macchiavello, Phys. Rev. Lett. {\bf 119}, 100502 (2017). 

\bibitem{schumachernielsen}
B. W. Schumacher and M. A. Nielsen, 
Phys. Rev. A \textbf{54}, 2629 (1996).

\bibitem{schumacher}
B. W. Schumacher, Phys. Rev. A \textbf{54}, 2614 (1996).



\bibitem{pla}  G. M. D'Ariano, P. Lo Presti, and M. F. Sacchi, Phys. Lett. A {\bf 272}, 32 (2000).

\bibitem{moore}A. Ben-Israel and T. N. E. Greville, {\em Generalized inverses: Theory and applications} (New York, Springer, 2003).


\bibitem{faith2}G. M. D'Ariano and M. F. Sacchi, J. Opt. B {\bf 7}, S408 (2005).

\bibitem{NC00} M. A. Nielsen and I. L. Chuang, {\em Quantum Information and
  Communication} (Cambridge, Cambridge University Press, 2000).


\bibitem{bellobs}G. M. D'Ariano, P. Perinotti, and M. F. Sacchi, J. Opt. B 
{\bf 6}, S487 (2004).


\bibitem{div}D. P. Di Vincenzo, P. W. Shor,  and  J. A.   Smolin, Phys. Rev. A {\bf 57}, 
830 (1998).

\end{thebibliography}
\end{document}